\documentstyle[prb,twocolumn,aps]{revtex}
\begin{document}
\tolerance 50000
\draft

\title{\bf Persistent currents in interacting systems: role of the spin.}
\author{Georges Bouzerar\cite{byline1} and Didier Poilblanc\cite{byline2}}
\address{
Groupe de Physique Th\'eorique,
Laboratoire de Physique Quantique,\\
Universit\'{e} Paul Sabatier,
31062 Toulouse, France}

\twocolumn[
\date{December 19}
\maketitle
\widetext
\vspace*{-1.0truecm}

\begin{abstract}
\begin{center}
\parbox{14cm}{
Persistent currents flowing through disordered mesoscopic
rings threaded by a magnetic flux are investigated.
Models of fermions
with on-site interactions (Hubbard model)
or models of spinless fermions with nearest neighbor interactions
are considered on 2D cylinders with twisted boundary conditions
in one direction to account for a magnetic flux.
Self-consistent Hartree-Fock methods are used to treat the
electron-electron interaction beyond first order.
We show that
the second harmonic of the current (which is relevant in the diffusive regime)
is {\it strongly suppressed} by the interaction
in the case of spinless fermions
while it is {\it significantly enhanced} in the Hubbard model.
Our data also strongly suggest that the reduction
(increase) of this harmonic
is related to a strong increase (reduction) of the spacial fluctuations
of the charge density.
}
\end{center}
\end{abstract}
\pacs{
\hspace{1.9cm}
 }
]
\narrowtext

The observations of mesoscopic currents in very pure metallic
nano-structures was done in pioneering experiments \cite{Levy,Chandra,Mailly}.
In the first case the experiment dealt with the average current
of a system of $10^{7}$ disconnected rings in the diffusive regime while in the
second a single ring was used.
Although the existence of such persistent currents in small metallic rings
was predicted long ago \cite{history1,Buttiker} the magnitude
of the observed currents
is still a real challenge to theorists.
There is a general belief that the interaction plays a crucial
role in enhancing the current.
But, so far, the role of the interaction in disordered systems is still
unclear. Treating interaction and disorder on equal footings is a difficult
task. Previous work \cite{meso1d,meso1d2} has shown by
exact diagonalizations (ED) of small clusters that, for strictly 1D systems
of spinless fermions, the effect of a {\it repulsive} interaction is
to increase further the localization of the
electrons and hence to decrease the value of the current.
Using a Hartree-Fock approach, Kato et al.\cite{note_1d}
have obtained a qualitatively good agreement with the exact
calculations. On the other hand, Giamarchi et al. \cite{giamarchi} have
pointed out that, for the 1D Hubbard model, i.e. when spin
is included, the interaction enhances the persistent current.
In this case, the increase of the current is closely related to
the {\it decrease} of the spacial fluctuations of the charge density
or, equivalently, to the smoothing out of the charge density
as it occurs in the 1D Hubbard model whith repulsive interaction.
This emphasizes the important role of the spin in 1D systems.

In higher dimensions the role of the spin is still unclear.
First order calculations for which spin is irrelevant have shown that the
persistent currents are increased by the interactions \cite{Ambegaokar}.
More recently, Ramin et al. \cite{Ramin} have numerically shown
that the first order Hartree-Fock (HF) correction
to the second harmonic of the persistent current was in
agreement with the analytical treatment \cite{Ambegaokar}.
Hence, for both the spinless fermion model
with nearest-neighbor interactions and the Hubbard model,
the second harmonic is enhanced.
However, in this treatment it is found that
a nearest neighbour interaction tends to decrease
the value of the typical current
while a repulsive extended Hubbard interaction enhances it.
In some previous work, Exact Diagonalisations (ED)
calculations have been compared to a Self-consistent Hartree-Fock (SHF)
treatment of the interaction. For the small clusters ($4\times4$ clusters)
which could be handled we have found a good
agreement between the two sets of data \cite{article3}.
This direct comparison with the exact results
has therefore established some degree of reliability of the
self-consistent Hartree-Fock approximation at least in the diffusive regime.
In this paper, we use both HF and SHF treatments of the interaction
between particles which enables us to treat much larger systems.
The SHF treatment takes into account
high order terms in the interaction and, simultaneously, deals with
quantum interference effects due to the disorder somehow exactly.
It is important to note that this
method is different from the usual perturbative approach \cite{Ambegaokar}
where the corrections to
the current due to the interacting term are
calculated perturbatively.
In contrast to their approach,
our procedure includes a resummation of higher order terms
through a self-consistency
relation, which turns out to become essential at moderate interaction.
Nevertheless, an exact connection with some diagrammatic expension is
a tedious problem and this issue probably deserves further study.
Note that our study also applies in principle to single
or multi-ring experiments.
 From a theoretical point of vue, the difference simply relies
in the absence or presence of particle number fluctuations.
In the following, we shall assume that the number of electron
of the ring is fixed.

This paper is organized as follows: first, we compare, for
both spinless and Hubbard models, the first order correction (in the
interaction) and the SHF correction to the second harmonic of the
persistent currents.
We show that, in the case of spinless fermions,
the two methods stay in good agreement only for rather small
values of the interaction parameter but
a complete disagreement appears for moderate values.
In contrast, in the Hubbard case,
the agreement between the two approaches is rather good.
Secondly, we present evidences that this {\it decrease} ({\it increase})
of the persistent currents
is directly related to the {\it increase} ({\it decrease}) of
the spacial fluctuations of the charge density from site to site
as the interaction is switched on.
Thirdly, the effect of the electron repulsion on the second
harmonic is shown to increase with the
system size.

The Hamiltonian is defined on a $L\times L$ lattice with periodic boundary
conditions in one direction (e.g. x direction) and reads:
\begin{eqnarray}
{\cal H} ={\cal H}_{K} + {\cal H}_{int} + {\cal H}_{des}  .
\label{hamilt}
\end{eqnarray}
\noindent
${\cal H}_{K}=\sum_{\bf i,\bf j}\, t_{\bf ij}\, c_{\bf i}^\dagger
c_{\bf j}$ is the usual kinetic part
containing the flux dependance,
$t_{\bf ij}=\frac{1}{2}\exp{\{ i\frac{2\pi\Phi}{L}(i_x-j_x)\}}$ if $\bf i$ and
$\bf j$ are nearest neighbor sites and 0 otherwise.
${\cal H}_{int}$ is the interacting part
and ${\cal H}_{des}$ is the term
due to the disorder,
\begin{eqnarray}
{\cal H}_{des}=\sum_{\bf i}\,w_{\bf i }\,n_{\bf i}
\end{eqnarray}
\noindent
where $n_{\bf i}$ is the local density operator at site $\bf i$
and $w_{\bf i }$ are on-site energies chosen
randomly in $[-\frac{W}{2},\frac{W}{2}]$.
When spin is included, $n_{\bf i}=n_{\bf i \uparrow}\,+n_{\bf i \downarrow}$
and the kinetic term contains an additional sum over the spin indices.
In the spinless fermion case the electron repulsion is given by
\begin{eqnarray}
{\cal H}_{int}^{S}= \frac{1}{2}\sum_{\bf i,\bf j}\, V_{\bf ij}\, n_{\bf
i}\,n_{\bf j} ,
\end{eqnarray}
\noindent
where $V_{\bf i,j}$ of strength $V=|V_{\bf ij}|$ only connects
nearest neighbor sites (screened interaction).
Lastly, the Hubbard interaction is defined by
\begin{eqnarray}
{\cal H}_{int}^{H}= U\sum_{\bf i}\,n_{\bf i \uparrow}\,n_{\bf i \downarrow} .
\end{eqnarray}

In the SHF approximation, the interaction part of the spinless fermion
model reduces to,
\begin{eqnarray}
{\cal H}_{int}^{S} =
& &-\sum_{\bf i,\bf j}
\delta t_{\bf i \bf j}\,c^{\dagger}_{\bf i}\,  c_{\bf j}\,+
\sum_{\bf i}\delta w_{\bf i}\,n_{\bf i} \nonumber \\
& & -\frac{1}{2}\,\sum_{\bf i,\bf j}\,V_{\bf ij}\, (\big< n_{\bf i}\big>_{0}
\big< n_{\bf j}\big>_{0}\,-\left|\,\big<c^{\dagger}_{\bf j}\,c_{\bf
i}\big>_{0}\,\right|^{2})
\end{eqnarray}
\noindent
where $\big<  \big>_{0}$ stands for the expectation value in the ground-state
wavefunction. The expression above contains two terms:
an Hartree term proportional to
$\delta w_{\bf i}=\sum_{\bf j \ne \bf i }V_{\bf ij}\big< n_{\bf j}\big>_{0}$
which comes out as an extra on-site disorder potential and a Fock term
proportional to $\delta t_{\bf i \bf j}=V_{\bf ij}\big<
c^{\dagger}_{\bf j}\,c_{\bf i}\big>_{0}$ which is an extra hopping amplitude.
The quantities $\big< n_{\bf j}\big>_{0}$
and $\big< c^{\dagger}_{\bf j}\,c_{\bf i}\big>_{0}$ are calculated
self consistently so that the SHF Hamiltonian itself depends on the
filling factor, the desorder strength etc...
Similarly, in this approximation ${\cal H}_{int}^{H}$ becomes,
\begin{eqnarray}
{\cal H}_{int}^{H}=U\sum_{\bf i} (\big< n_{\bf i \uparrow}\big>_{0}\,
 n_{\bf i \downarrow}+\big< n_{\bf i \downarrow}\big>_{0}\,n_{\bf i \uparrow}
-\big< n_{\bf i \downarrow}\big>_{0}\big< n_{\bf i \uparrow}\big>_{0}) .
\end{eqnarray}
\noindent
Since we are interested in the paramagnetic phase we shall assume
$\big< n_{\bf i \downarrow}\big>_{0}=\big< n_{\bf i \uparrow}\big>_{0}$.
In this case the spin $\uparrow$ and $\downarrow$ are decoupled so that one can
write ${\cal H}_{int}^{H}=\sum _{\sigma}{\cal H}_{int}^{\sigma}$ where,
\begin{eqnarray}
{\cal H}_{int}^{\sigma}=U\sum_{\bf i} \big< n_{\bf i \sigma}\big>_{0} n_{\bf i
\sigma}
-\frac{1}{2} \big< n_{\bf i \sigma}\big>_{0}^{2} .
\end{eqnarray}

The current is defined as the derivative of the total energy
versus flux,
\begin{eqnarray}
I(\Phi) =
-\frac{\partial E(\Phi)}{\partial \Phi}
\end{eqnarray}
where $E(\Phi)$ is the total energy.
The current is a periodic function of $\Phi$ of period 1
($\Phi$ is measured in units of $\Phi_{0}$) and
thus can be expanded as a Fourier series,
\begin{eqnarray}
I(\Phi) =
\sum_n
I_{hn}\,sin(2\pi\,n\Phi)
\label{Ih}
\end{eqnarray}
where $I_{hn}$ are the harmonics of the current.
It is now well established that the ensemble average (over filling or disorder)
suppresses the first harmonic of the current \cite{gilles2}
\cite {article3}. This fact was indeed observed in multi-ring experiments
where the current was found to be $\frac{1}{2}$ periodic.
Therefore, in the following we shall focus on the second harmonic $I_{h2}$.
The solution of the previous set of non-linear self-consistent equations are
obtained
numerically on small clusters by an iterative procedure for arbitrary
values of $\Phi$ and arbitrary disorder realisations. The various physical
quantities are then averaged over disorder. The disorder average denoted
by $\big< \big>_{dis}$ in the following corresponds typically to
an average over at least 1000 configurations of the disorder.

As noted previously, the self-consistency relations can take care of
high order effects in the interaction. It is therefore necessary,
as a preliminary study, to investigate the role of the self-consistency
by comparing the SHF results to the simple first order calculations (refered to
hereafter as HF). In Fig. \ref{f1ab}(a) $\big< I_{h2} \big>_{dis}$
for spinless particles is plotted as a function of V.
The calculations have been done at half-filling on a $8 \times 8$ cylinder.
The HF contribution is of course linear in V.
As expected, we observe for small values of the interaction
a perfect agreement between the SHF and the HF calculations.
Indeed, the slope at V=0 of the SHF results
is actually given by the HF data.
However, for increasing V the SHF calculation shows a strong reduction
of the current whilst the HF predicts an increase.
We also observe that the region of agreement between HF and SHF
is reduced as the disorder increases.
In other words, this means that, as the strength of the interaction increases,
the effect of higher order terms becomes more and more dominant
and thus a first order calculation is not sufficient.
We will see later on that this reduction is
related, as in the 1D case, to an increase of the spacial fluctuations of the
charge density. The influence of the filling factor and of the
disorder strength on $\big< I_{h2} \big>_{dis}$
vs V is shown in Fig. \ref{f1ab}(b).
This figure shows that the repulsive interaction is detrimental
to the persistent currents which are drastically suppressed for a wide
range of parameters. Note that, at quarter filling, $\big< I_{h2} \big>_{dis}$
starts immediately to decrease with V contrary to the behavior observed
at half-filling.

In contrast to the spinless model, the Hubbard model exhibits a completely
different behavior as seen in Fig. \ref{f2} showing the relative increase of
$\big< I_{h2} \big>_{dis}$ as a function of the Hubbard repulsion U.
Interestingly enough, both SHF and HF calculations predict an increase of
the current. It is also interesting to notice that the first order calculation
gives reliable results regarding the effect of the interaction.
However, the first order calculations always predict
larger values of the currents.

At this point, these results already suggest that the nature
of the interaction plays a crutial role: in the spinless case the currents
are strongly reduced by the interaction in contrast to a
Hubbard repulsive interaction which enhances the currents.
Secondly, we have shown that, in the spinless fermions case,
higher order terms become rapidly dominant even for relatively small
values of the interaction strength.
Hence a first order approach is not sufficient.

Let us now try to developp a physical picture that could help
to understand the role of the spin.
We shall argue that the enhancement (reduction) of the persistent
currents is related to the reduction (increase) of the
fluctuations of the charge density $n_{\bf i}$ from site to site.
One way of observing this effect on the charge density
consists in plotting the distribution of the
local density $\big< n_{\bf i}\big>_{0}$.
For that purpose, we shall consider here a $10 \times 10$ system
and assume that the $\big< n_{i} \big>_{0}^{k}$ (where
the subscript k labels the various realisations of the disorder) are
independant variables.
The distribution of the charge densities can then be defined as,
\begin{eqnarray}
P(\rho)=\frac{1}{N_{dis}L^2} \sum_{k=1}^{N_{dis}}\sum_{\bf i}
\delta(\big< n_{\bf i} \big>_{0}^{k} - \rho)
\end{eqnarray}
where L is the length of the system (L=10) and $N_{dis}$ is
the number of disorder configurations.
As usual, the local densities $\big< n_{i}\big>^{k}_{0}$ are calculated
self-consistently. Note that, for a sufficiently large system, we expect
the distribution to become independant of the choice of the
disorder configurations.

For spinless fermions, we have plotted in Figs. \ref{f3ab}(a,b)
the distribution $P(\rho)$ calculated on a
$10 \times 10$ cylinder for two different fillings.
We clearly observe that, with the interaction, both the shape of
the distribution change and the width of the distribution increases.
These effects are strongly emphasized in Fig. \ref{f3ab}(b) in which two peaks
appear, showing a tendancy towards the formation of a charge density wave
as we approach commensurability (n=0.5).
Note that such an instability is clearly unphysical for a metal
and could be easily removed by preventing perfect nesting of the Fermi surface
(for example, by including hopping terms at larger distances).
However, the effect of the interaction
to broaden the distribution $P(\rho)$ seems to be independant of the precise
details of the band structure and is generic for any filling.
In contrast, in the case of the Hubbard model (Fig. \ref{f4}), we observe
the opposite effect: the distribution
shrinks around the average value as the interaction is switched on.

Let us now turn to more qualitative
results. The width of the distribution (assuming independant
variables for different configurations of the disorder) $\delta \rho $
is given by
\begin{eqnarray}
\delta \rho=\frac{1}{N_{dis}} \sum_{k=1}^{N_{dis}}
\sqrt{\frac{1}{L^{2}} \sum_{\bf i}
(\big< n_{\bf i} \big>_{0}^{k} - n)^{2}} .
\end{eqnarray}
\noindent where $n=\frac{N_{e}}{L^{2}}$
is the filling ($N_{e}$ is the number of electrons).
We have studied the effect of the interaction on $\delta \rho $
as a function of V (spinless fermions) or U (Hubbard model)
for different values
of the disorder strength W and different fillings.
In Fig. \ref{f5ab}(a) $\delta \rho (V,W)$ is plotted
at half-filling as a function of the parameter V (spinless fermions).
Here we clearly observe an increase of the width $\delta \rho$ of
the distribution for increasing V. Very crudely, this effect seems
to depend only on the combined parameter $\frac{V}{W}$.
In contrast, in the Hubbard case shown in Fig. \ref{f5ab}(b), we observe
a reduction of $\delta \rho (U,W)$ as U increases.
In Figs. \ref{f6ab}(a,b) we have plotted the relative variations of
the width of the distribution
$\delta\rho(V,W)/\delta\rho(0,W)$ and $\delta\rho(U,W)/\delta\rho(0,W)$
as a function of the interaction parameters V and U for various densities.
We observe that the effects described above become stronger
for larger fillings i.e. when the interaction between the particles becomes
more effective. Note however that, in the case of the spinless fermion
model, this behavior is unrelated to the commensurability
since densities n=0.25 and n=0.4 have been chosen, clearly away
from half-filling (n=0.5).
In Figs. \ref{f6ab} we see that, in the spinless case, the
width has increased by almost a factor 4 for $V=0.8$ and $N_{e}=40$.
In the Hubbard case the reduction of
$\delta \rho(U,W)$ is 25\% larger compared to half-filling.
Our present study then strongly suggests that a reliable explaination of
the observed large persistent currents must somehow take into account
the spin of the particles.

We finish our discussion by a study of the influence of the system size
on our results.
In Fig. \ref{f7} we have plotted the relative increase of the current
$\big< I_{h2}\big>_{dis} (U,W)/\big< I_{h2}\big>_{dis} (0,W)$
vs U at fixed density but for different system sizes.
This figure clearly indicates that, as the size of the system
increases, the effects induced by the interaction become stronger.
We also expect that this is also true when the connectivity of the
lattice increases (i.e. going form 2D to real 3D rings).
Although a systematic accurate study as a function of the
sample size is not feasible our data for the Hubbard model for,
let say U=0.4, are not
inconsistent with the magnitude of the currents observed in the experiments.

In conclusion, we have shown that, in the spinless fermion model,
the effect of a moderate
interaction leads to a drastic reduction of the magnitude of the current,
contrary to what is expected from first order calculations.
For the first time, we have established that, in 2D, taken into account
the spin degrees of freedom is
crucial to explain the enhancement of the current due to the interaction,
similarly to the 1D case.
Although the calculations presented in this paper deal with
rather small 2D clusters, we expect that, for larger systems
and higher connectivity (e.g. in 3D) and in the presence
of an Hubbard interaction, the impurity scattering
will become even less effective to localize the electrons.
Although a systematic study as a function of the
sample size is still out of reach of present day computors,
our data of the Hubbard model at intermediate U (around 0.4) are not
inconsistent with the magnitude of the current observed in
the experiments. Our study suggests that the SHF approach provides
a relatively good tool to study, on equal footings, the effects of
the interaction and of the disorder.

We gratefully acknowledge stimulating discussions with T. Giamarchi
and G. Montambaux. D.P. acknowledges support from the
EEC Human Capital and Mobility Program under grant CHRX-CT93-0332.
{\it Laboratoire de Physique Quantique (Toulouse)}
is {\it Unit\'e Associ\'ee No. URA505 du CNRS}.

%
%
\begin{figure}
\caption{$\big< I_{h2} \big>_{dis}$ as a function of V
calculated within the SHF method for spinless fermions on
a $8 \times 8$ cylinder. An average over 1000 disorder
configurations has been done.
(a) Comparison between first order HF (open symbols)
and SHF results (full symbols) at half-filling.
Circles correspond to $W=3$ and squares to $W=4$.
(b) SHF results at half-filling (full symbols) and quarter filling
(open symbols).
Circles correspond to $W=3$, squares to $W=4$ and
triangles to $W=5$.
}
\label{f1ab}
\end{figure}

%
%
\begin{figure}
\caption{Ratio $\big< I_{h2} \big>_{dis}(U,W)/\big< I_{h2} \big>_{dis}(0,W)$
as a function of U calculated in the Hubbard model
within both HF (open symbols)and SHF (full symbols) methods on
a $8 \times 8$ cylinder. An average over 1000 disorder realisations
have been performed. Circles correspond to $W=3$ at quarter-filling,
squares to $W=3$ at half-filling and triangles
to $W=4$ at half-filling.
}
\label{f2}
\end{figure}

%
%
\begin{figure}
\caption{Distribution $P(\rho)$ in the case of spinless
fermions.
The calculations have been done at quarter filling
$n=0.25$ (a) and $n=0.4$ (b) on a $10 \times 10$ cylinder.
We have averaged over 30 configurations of the disorder.
The values of V and W are given in the figure.
}
\label{f3ab}
\end{figure}

%
%
\begin{figure}
\caption{Distribution P($\rho$) in the Hubbard case.
The calculations have been done at half-filling ($n=1$)
on a $10 \times 10$ cylinder.
We have considered 30 configurations of the disorder.
The values of U and W are given in the figure.
}
\label{f4}
\end{figure}

%
%
\begin{figure}
\caption{Effect of the interaction on the width $\delta \rho
(V,W)$ at half-filling as a function of the interaction
parameter (U or V). The calculations have been done on a
$10 \times 10$ cylinder and 30 configurations of the disorder
have been considered. The values of W are given in the figure.
(a) Spinless fermion model; (b) Hubbard model. }
\label{f5ab}
\end{figure}

%
%
\begin{figure}
\caption{Ratio $\delta \rho (V,W)/\delta \rho (0,W)$ as a function
of the interaction parameter (U or V) for different filling factors
(the filling is indicated on the plot)
and values of W.
The calculations have been done on a $10 \times 10$ cylinder
with 30 configurations of the disorder.
(a) Spinless fermion model ($n=0.25$ and $n=0.5$); (b) Hubbard model
($n=0.5$ and $n=1$). }
\label{f6ab}
\end{figure}

%
%
\begin{figure}
\caption{Ratio $\big< I_{h2} \big>(U,W)_{dis}/\big< I_{h2} \big>(0,W)_{dis}$
as a function of U (Hubbard
case) calculated within the SHF method.
The calculations have been done on $6\times 6$ (circles),
$8\times 8$ (squares) and $10\times 10$ (triangles) cylinders at
half-filling ($n=1$) and for $W=3$. 1000 realisations of the disorder have been
used.
}
\label{f7}
\end{figure}

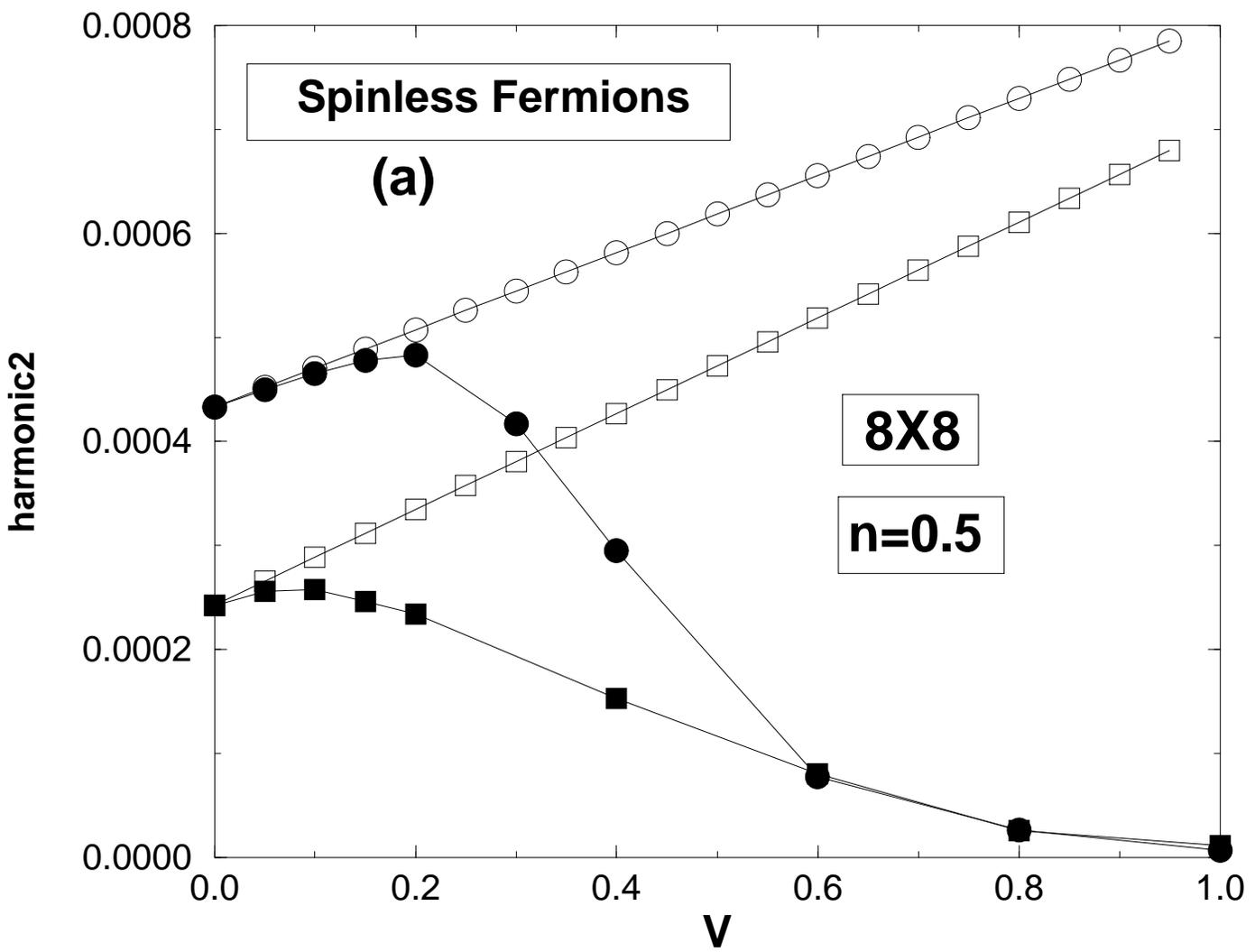
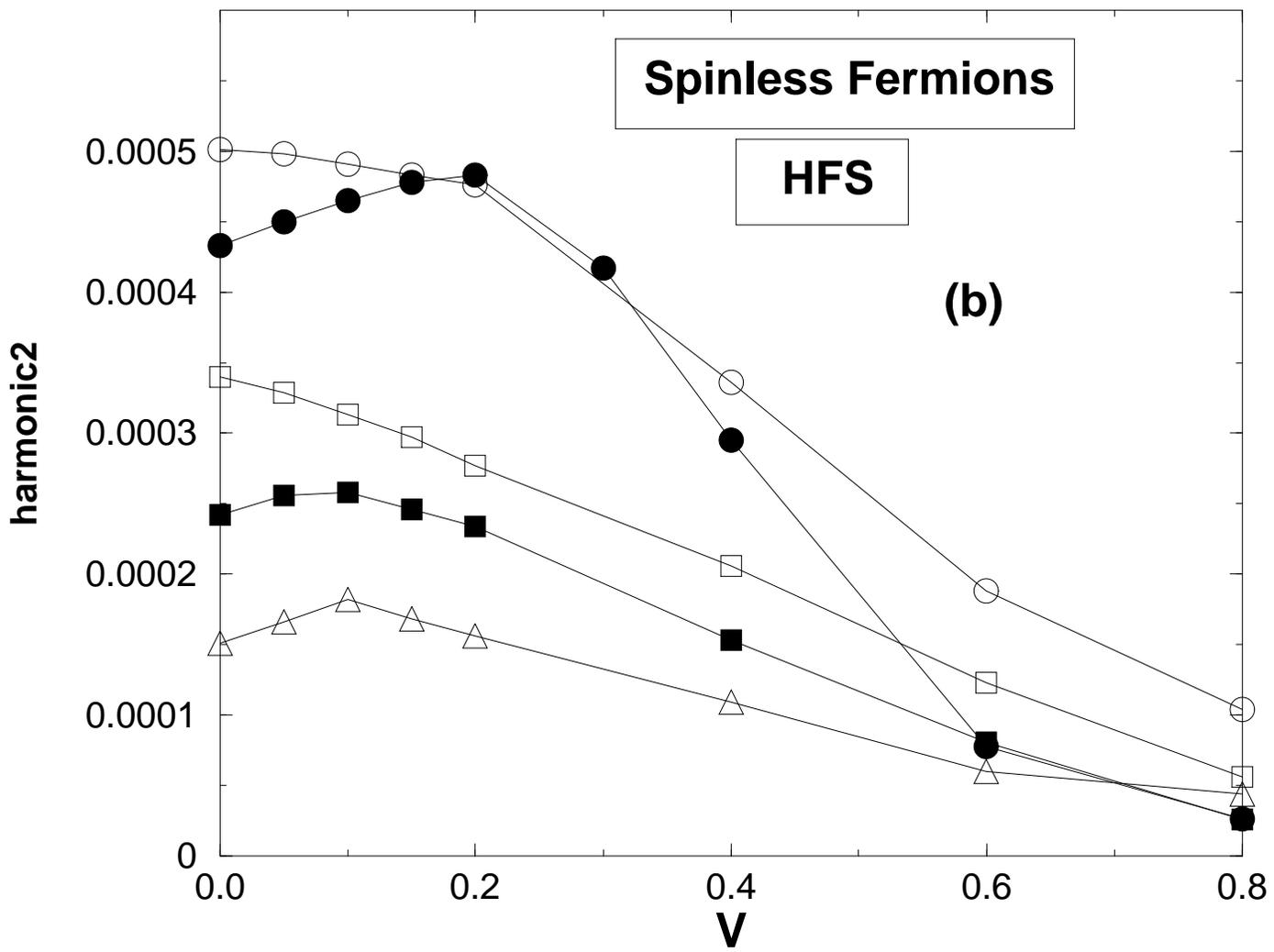
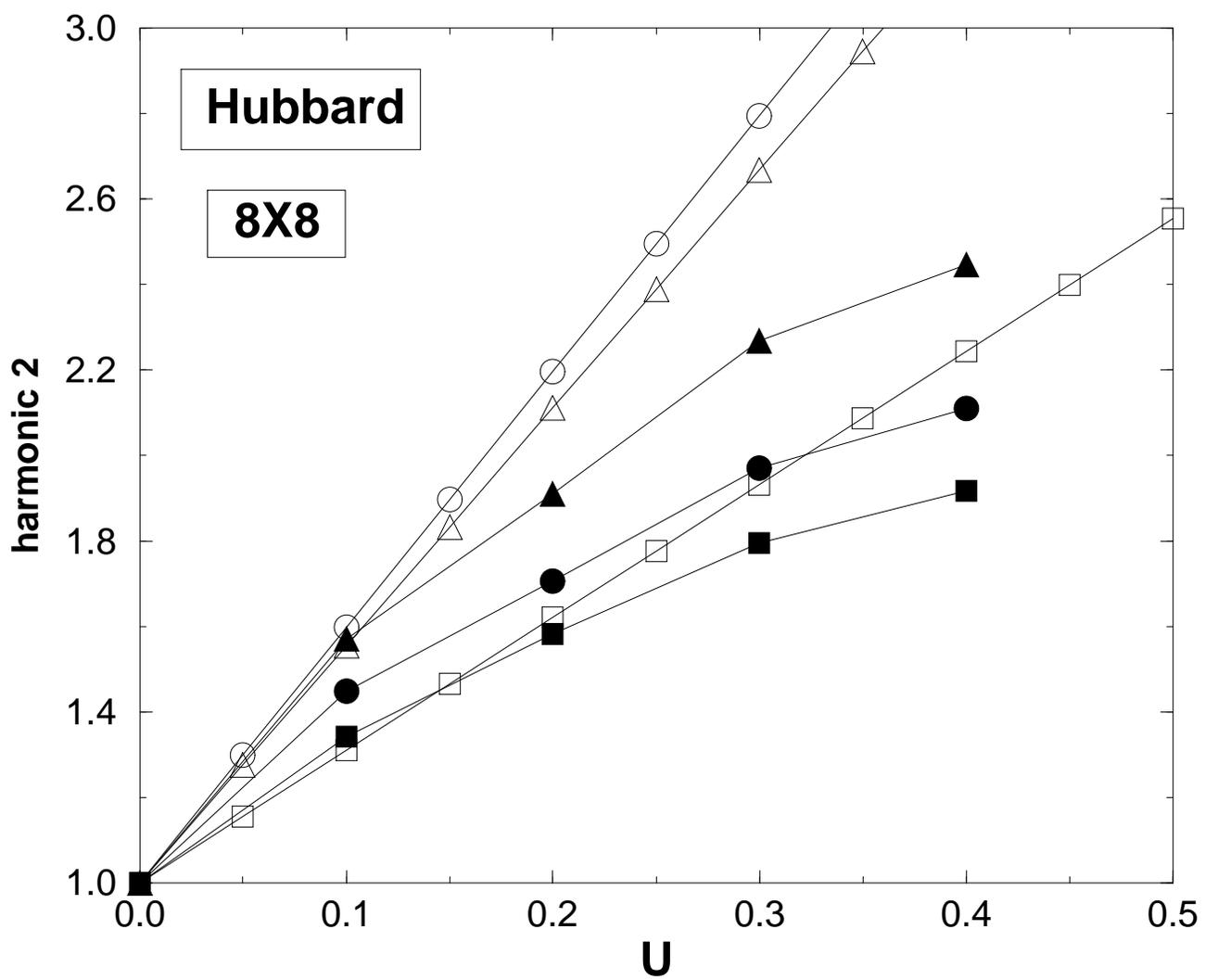
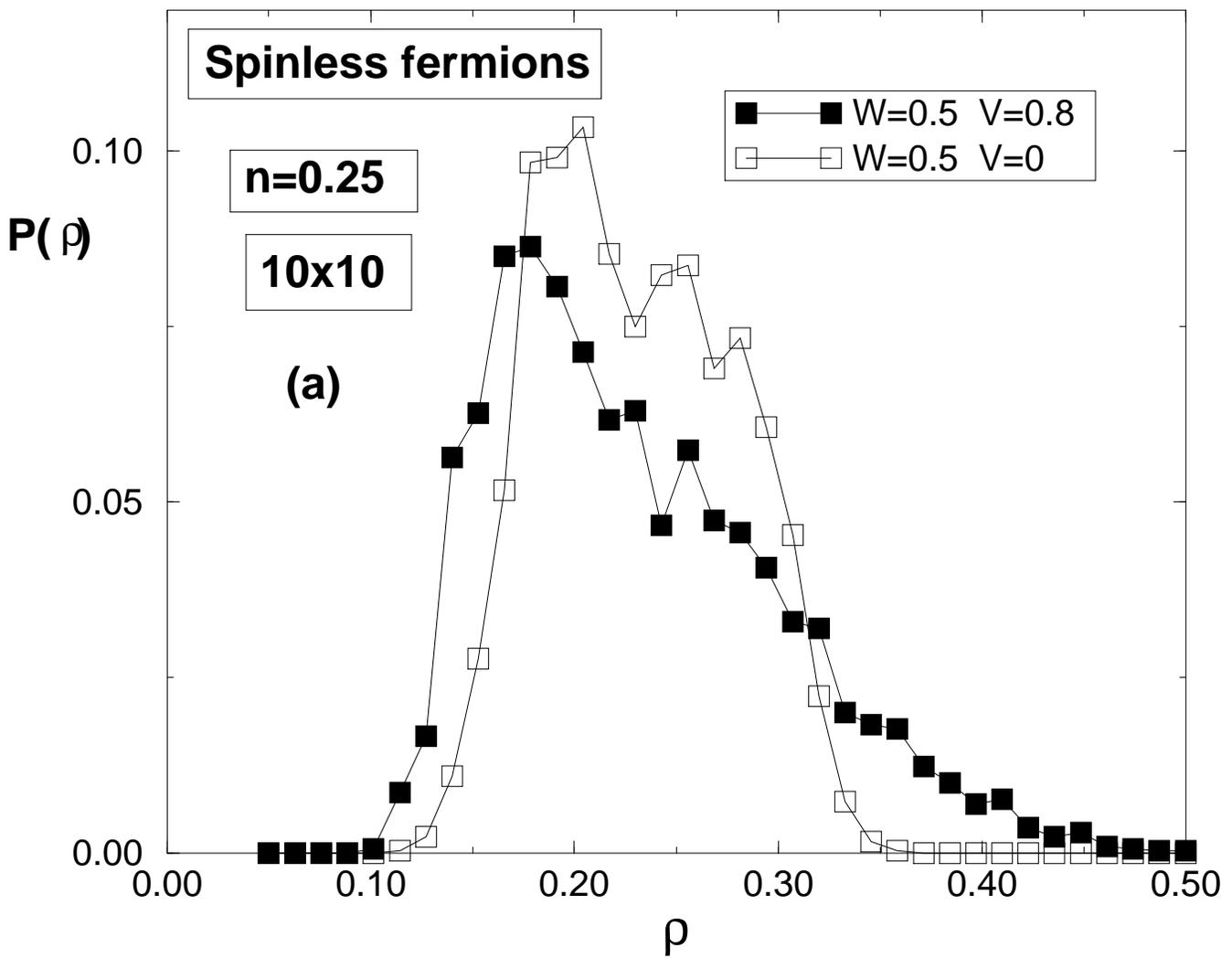
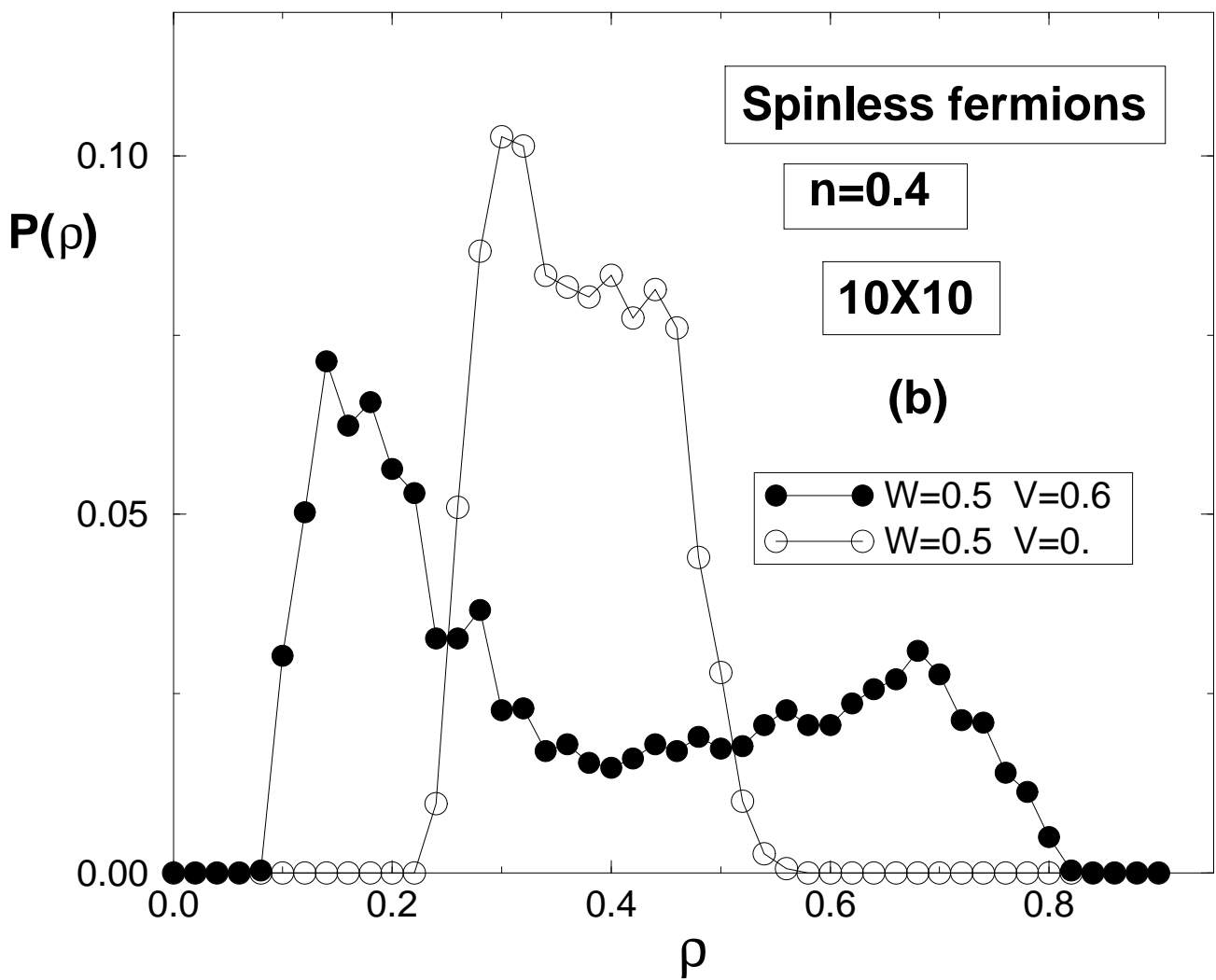
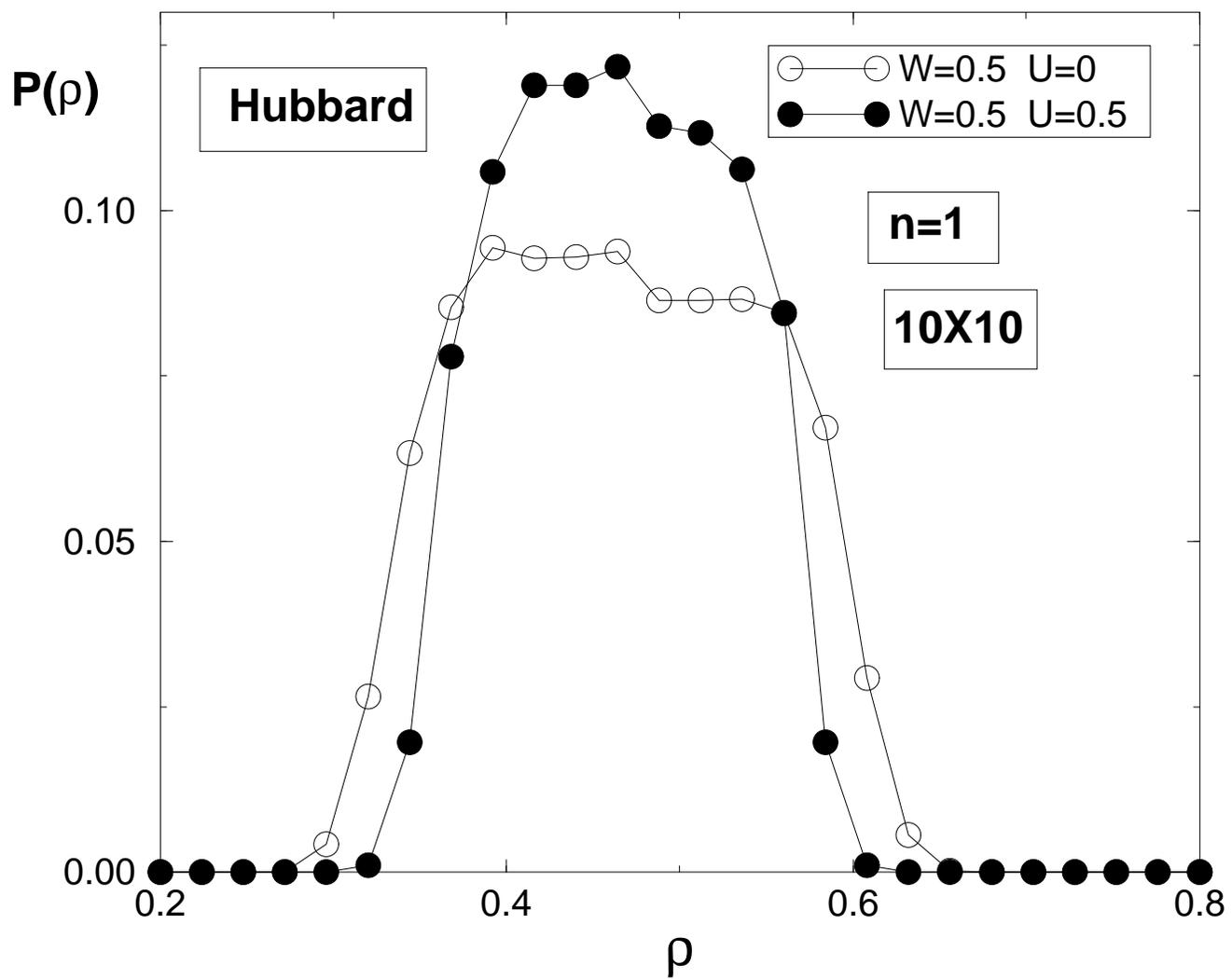
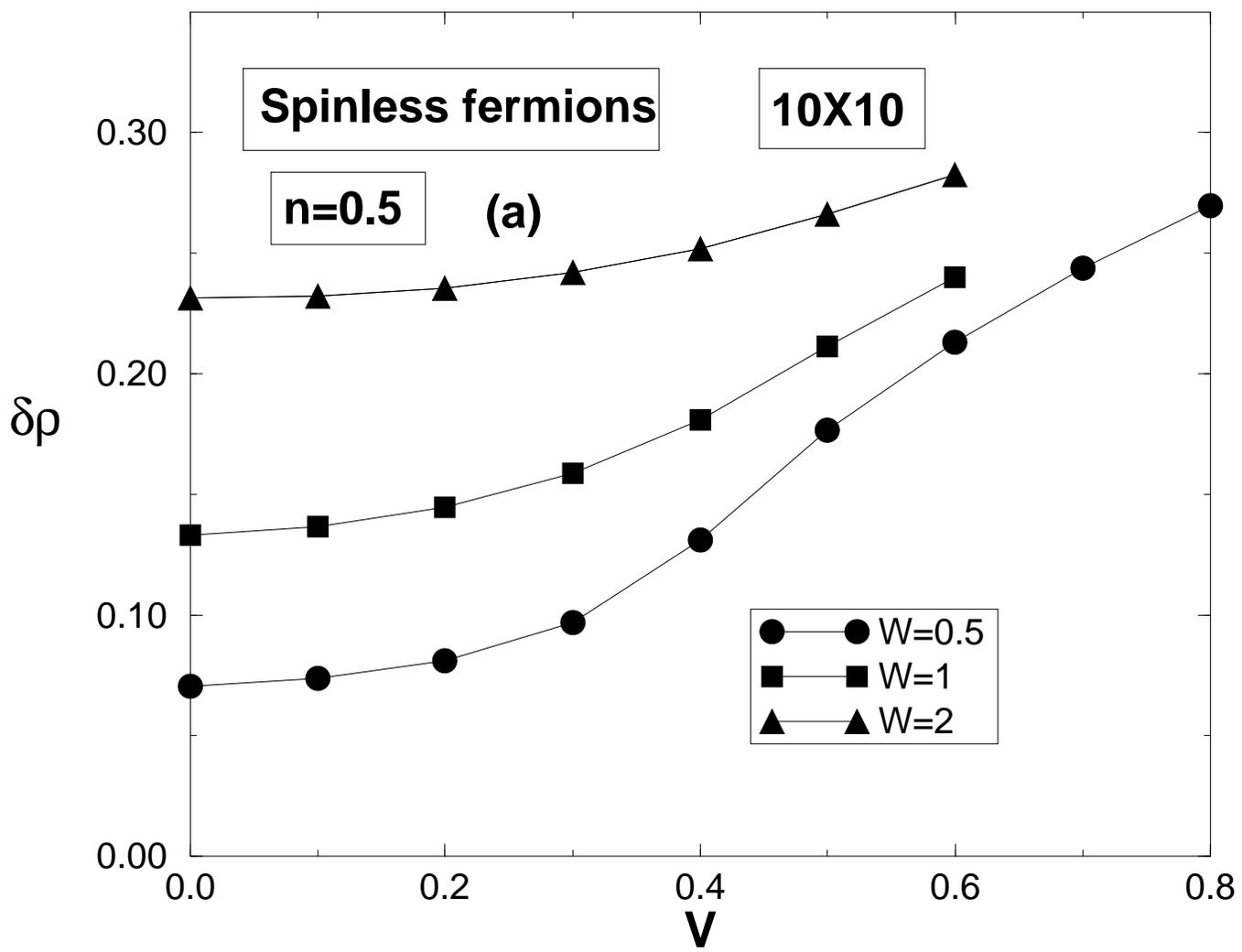
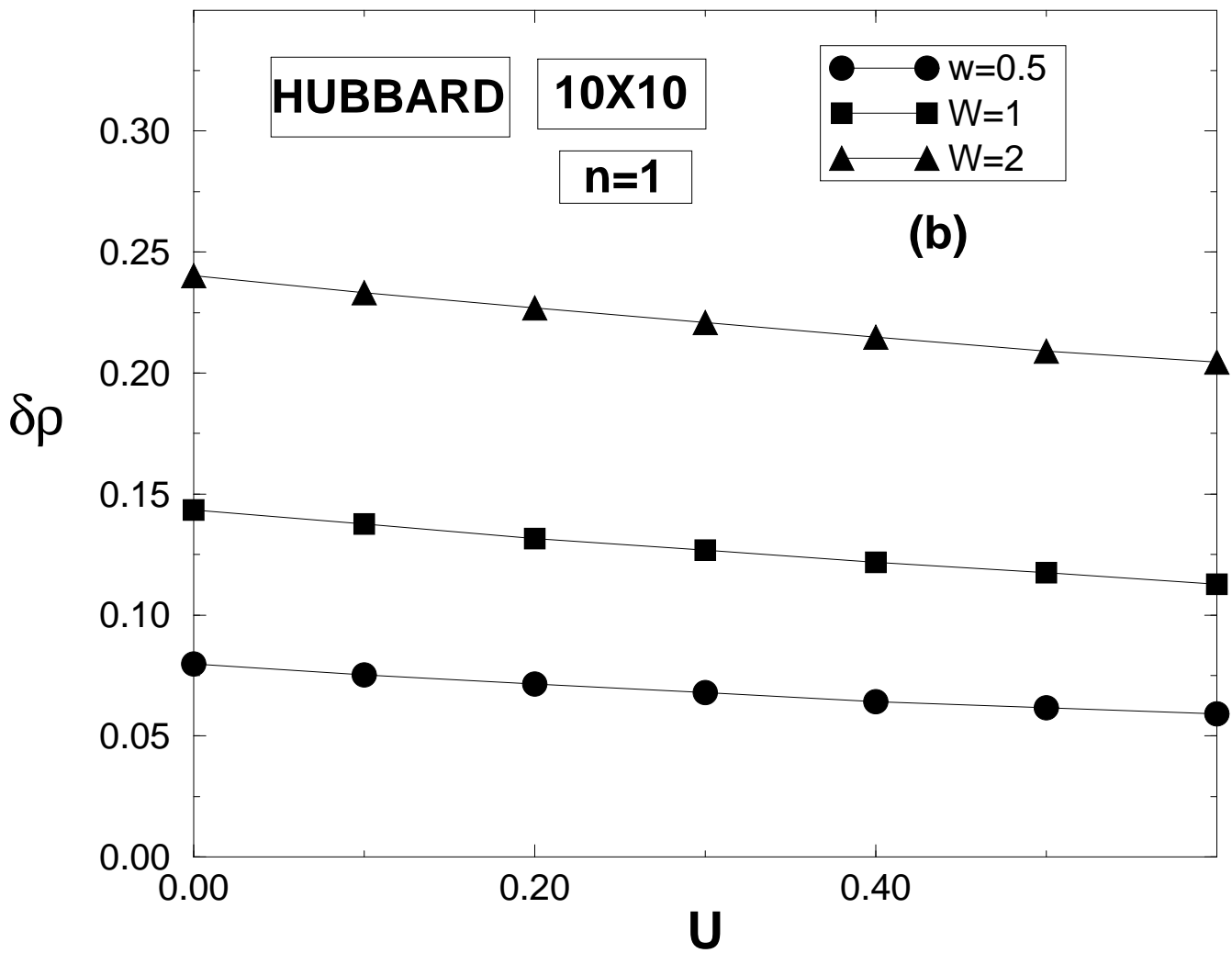
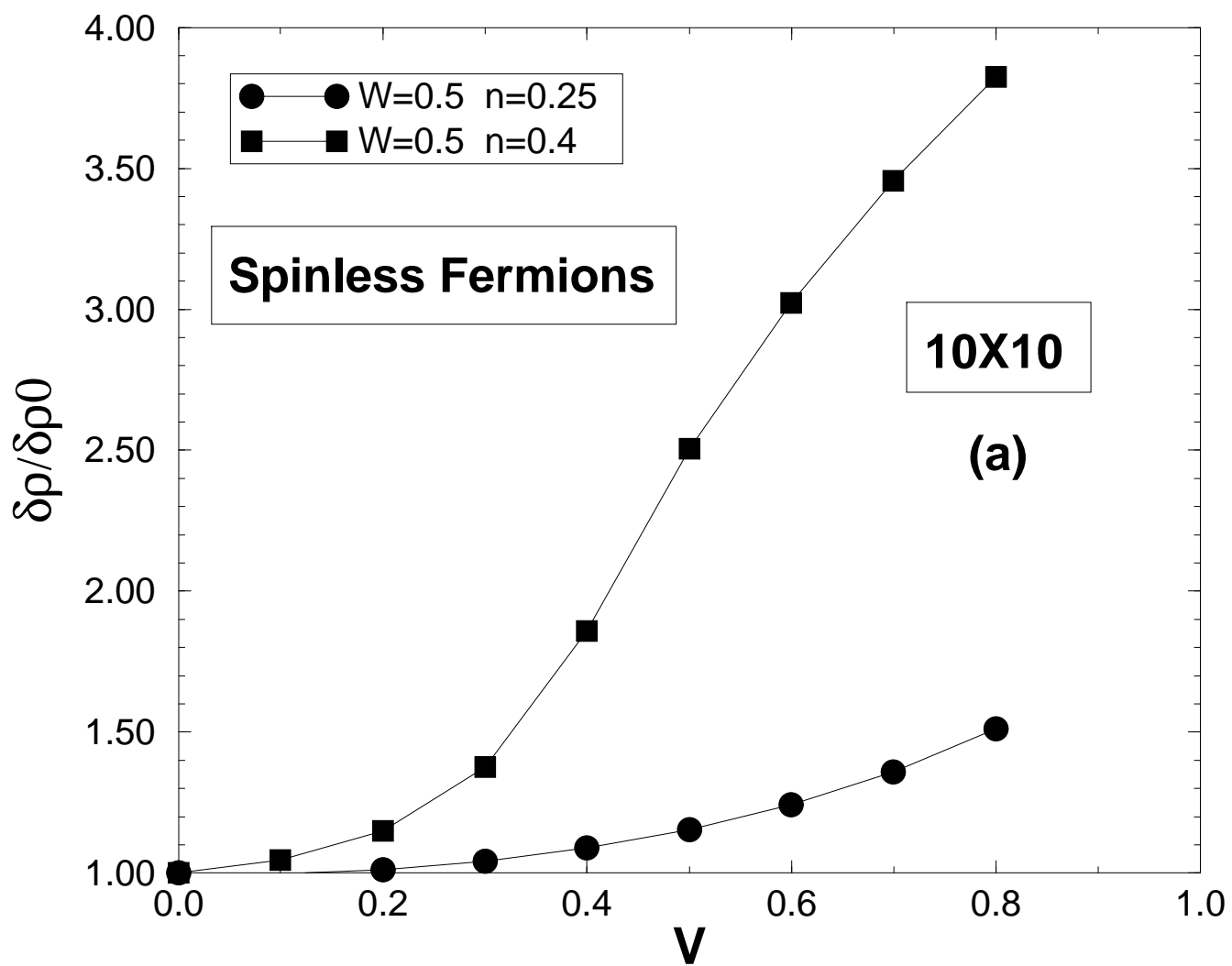
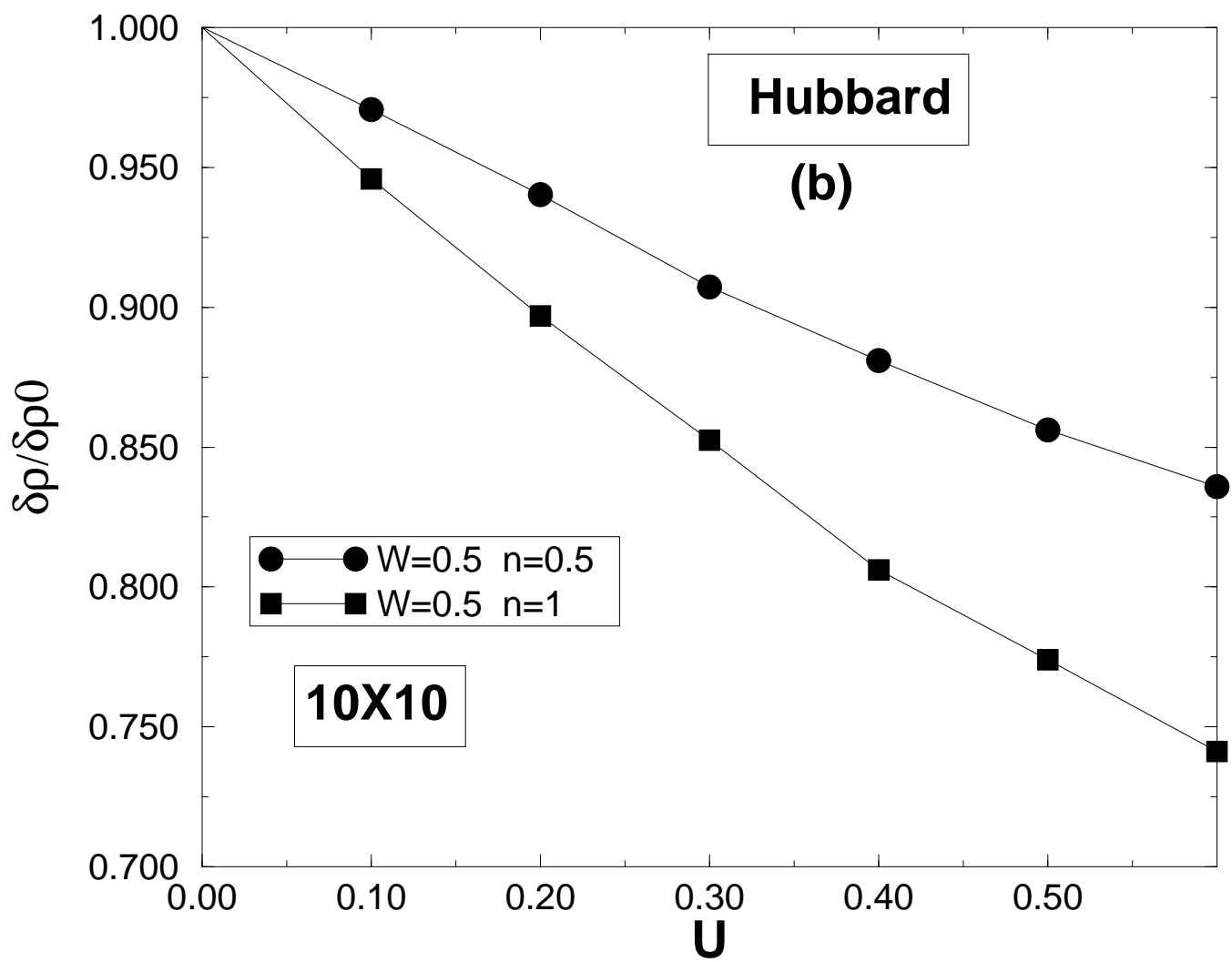
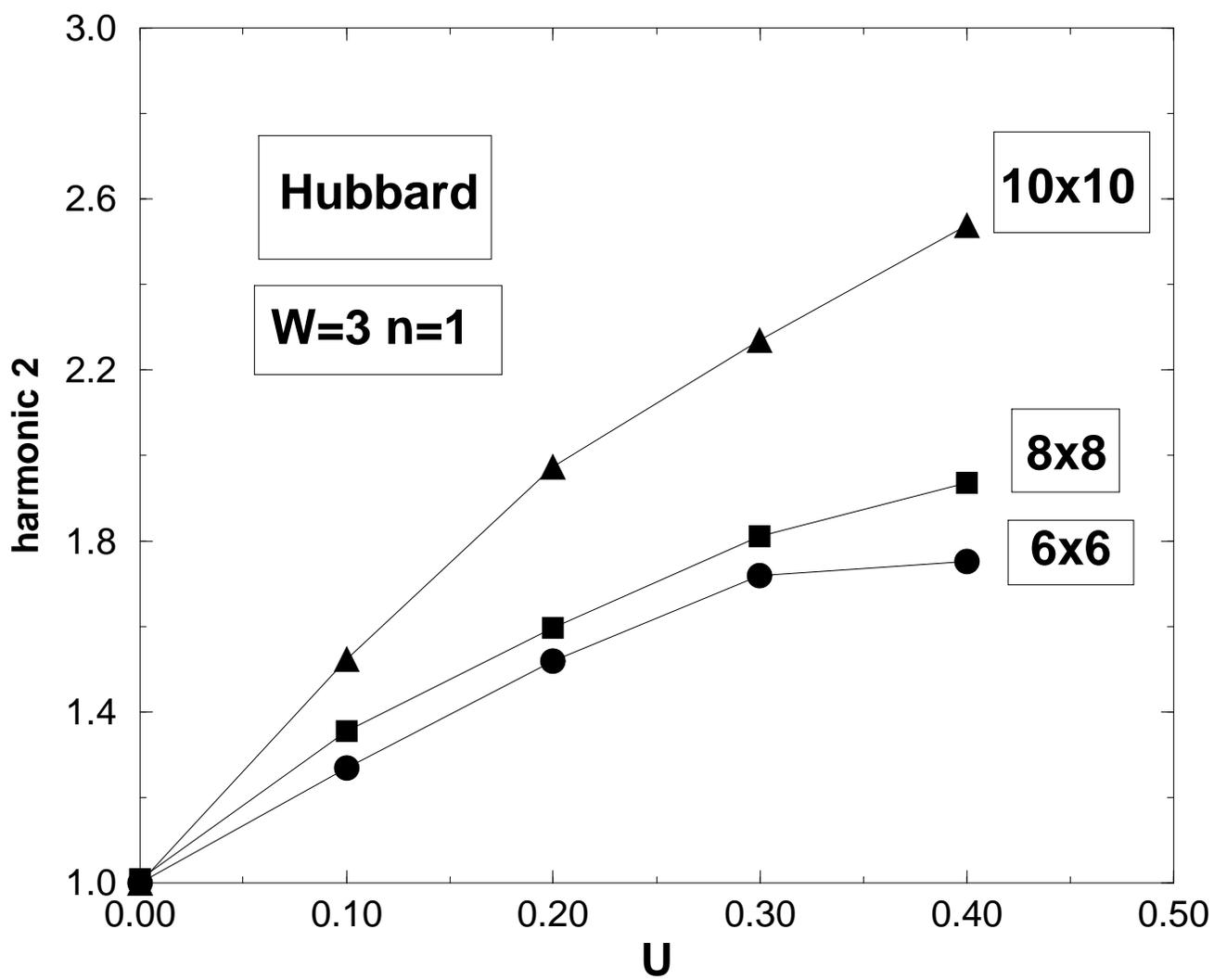
\end{document}